\documentclass[aps,prl,superscriptaddress,showpacs,twocolumn]{revtex4}
\usepackage{units}
\usepackage{amsmath}
\usepackage{amssymb}
\usepackage{graphicx}
\usepackage{bm,color,microtype}

\newcommand{\bracket}[1]{\langle#1\rangle}

\newcommand{\be}{\begin{equation}}
\newcommand{\ee}{\end{equation}}
\newcommand{\bea}{\begin{eqnarray}}
\newcommand{\eea}{\end{eqnarray}}

\begin{document}
\title{Pattern Formation and Strong Nonlinear Interactions in Exciton-Polariton Condensates}

\author{Li Ge}
\affiliation{\textls[-18]{Department of Engineering Science and Physics, College of Staten Island, CUNY, New York 10314, USA}}
\affiliation{Department of Electrical Engineering, Princeton University, Princeton, New Jersey 08544, USA}
\author{Ani Nersisyan}
\author{Baris Oztop}
\author{Hakan E. T\"ureci}
\affiliation{Department of Electrical Engineering, Princeton University, Princeton, New Jersey 08544, USA}

\date{\today}

\begin{abstract}
Exciton-polaritons generated by light-induced potentials can spontaneously condense into macroscopic quantum states that display nontrivial spatial and temporal density modulation. While these patterns and their dynamics can be reproduced through the solution of the generalized Gross-Pitaevskii equation, a predictive theory of their thresholds, oscillation frequencies, and multi-pattern interactions has so far been lacking. Here we represent such an approach based on the linear non-Hermitian modes of the complex-valued light-induced potential. We provide a simple analytic expression for the lowest thresholds that is able to explain the modal patterns observed in recent experiments for various pump geometries. We also show that the evolution of the condensate with increasing pump strength is strongly geometry dependent and can display contrasting features such as enhancement or reduction of the spatial localization of the condensate. 
\end{abstract}

\pacs{71.36.+c,67.85.Hj,42.65.Pc,05.30.Jp}
\maketitle

Realizing and visualizing macroscopic quantum states has attracted considerable interest in the past decades. Using exciton-polaritons, quasi-particles formed by strong light-matter interactions in an optical cavity, this goal has been achieved in non-equilibrium systems \cite{rmp1,Snoke,rmp2}. A particularly promising method to generate polariton condensates is through  non-resonant optical pumping of a semiconductor microcavity \cite{Richard1,Richard2,Kasprzak,Christ,Balili,Baas,Bloch1,Bloch2,Manni,Assman,Baumberg1,Baumberg1.5,Baumberg2,nonresonant2,nonresonant3,Bloch3}. Injected incoherent carriers rapidly cool and scatter into localized polariton condensates above successive thresholds. Such all-optical spontaneous patterning of condensates presents a versatile method to prepare and manipulate polariton condensates, critical to realization of extended polaritonic circuits \cite{Bloch3,Shelykh, Malpuech1, Malpuech2, Liew1,Liew2}.

Earlier experimental work \cite{Bloch2,Manni,Baumberg1,Baumberg1.5,Baumberg2,nonresonant2,nonresonant3} has demonstrated that the spatio-temporal patterns formed strongly depend on the geometry of the pump spot in a nontrivial manner. Much of the phenomenology of pattern formation can be reproduced through a theoretical model based on the generalized Gross-Pitaevskii equation (GPE) \cite{Wouter_prl07,Wouter_prb08,Sarchi_prb08}, demonstrating that the patterns are a result of the complex interplay of generation and leakage of polaritonic quasi-particles, as well as the nonlinear interactions between them. While the observed phenomena are similar to multi-mode lasing, two striking differences are notable. Firstly, polaritonic condensates display a sizable $\chi^{(3)}$-nonlinearity. Secondly, the polaritonic medium is translationally invariant in the absence of the pump (in the plane of the condensate), in contrast to a standard laser cavity which is characterized by long lived spatially localized resonances \cite{bibnote:1}. As such, a predictive theory for polaritonic pattern formation based on linear resonances has so far been lacking.

In this letter, we present a theory of pattern formation in optically generated potentials that builds on non-Hermitian, current-carrying quasi-modes. The presented theory allows us to capture the patterns formed in the steady-state directly, accounting for nonlinearities exactly, instead of going through time-dependent simulations of the GPE. We find a simple but powerful expression for thresholds of condensation and the associated frequencies of oscillations, quantifying the contribution of particle formation, leakage, and interactions. Our findings indicate the possibility of having either an isomorphically or non-isomorphically trapped pattern with the lowest threshold, distinguished by whether the condensate pattern follows the geometry of the pump.
We also discuss the evolution of the condensate with the pump strength, which is highly geometry dependent and can display contrasting features such as enhancement or reduction of the spatial localization of the condensate.

We consider a planar geometry where the condensate is confined by two mirrors facing each other. The injected incoherent carriers act both as sources and as a repulsive potential for condensed polaritons, captured by a non-Hermitian effective Hamiltonian \cite{Wouter_prl07}
\be
H(\vec{r},t) = -\frac{\nabla^2}{2m} + V(\vec{r},t) + \frac{i}{2}\left[R n_{\!\scriptscriptstyle R}(\vec{r},t)-\gamma_c\right].\label{eq:GPE0}
\ee
Here $m$ is the effective mass of the lower polariton branch, $n_{\!\scriptscriptstyle R}(\vec{r},t)$ is the density of the exciton reservoir, and we have taken $\hbar=1$.
$\gamma_c$ is the mirror loss and the term $R n_{\!\scriptscriptstyle R}(\vec{r},t)/2$ represents the generation rate of condensed polaritons. When the relaxation of $n_{\!\scriptscriptstyle R}(\vec{r},t)$ is fast \cite{Wouter_prl07},
its temporal dependence can be eliminated adiabatically, i.e. $n_{\!\scriptscriptstyle R}(\vec{r})\approx Pf(\vec{r})[\gamma_{\!\scriptscriptstyle R}+R|\Psi(\vec{r},t)|^2]^{-1}$, where $\gamma_{\!\scriptscriptstyle R}$ is the decay rate of the exciton reservoir, $\Psi(\vec{r},t)$ is the condensate wave function, and $P$ and $f(\vec{r})$ are the {\it area-averaged} strength and the spatial profile of the pump, with $\int f(\vec{r}) d\vec{r}$ equals the area of the pump region.
The effective potential $V(\vec{r},t) = g_{\!\scriptscriptstyle R} n_{\!\scriptscriptstyle R}(\vec{r},t) + g|\Psi(\vec{r},t)|^2$ describes the repulsive polariton-exciton reservoir and polariton-polariton interactions.
We assume that the condensate can be described by a steady state of the multi-periodic form
\be
\Psi(\vec{r},t) = \sum_\mu \psi_\mu(\vec{r};P) e^{-i\omega_\mu(P) t},\label{eq:multiperiodic}
\ee
where $\omega_\mu(P)$ is the frequency of each condensate pattern $\psi_\mu(\vec{r};P)$, formed above their respective threshold $P_\mu$.

Next we introduce the {\it linear} threshold $P^{(0)}_\mu$ for condensation into a particular pattern $\psi_\mu$, which does not take into account its nonlinear interactions with other patterns. The lowest linear threshold $P^{(0)}_{\mu=1}$ is exact and provides $P_{\mu=1}$, which helps identifying the first condensate pattern formed as the pump increases. The higher-order ones $P^{(0)}_{\mu>1}$ differ from the actual thresholds, and the differences indicate the strength of the nonlinear interactions among the condensate patterns, either directly or via the density of the exciton reservoir.

\begin{figure}[t]
\begin{center}
\includegraphics[clip,width=\linewidth]{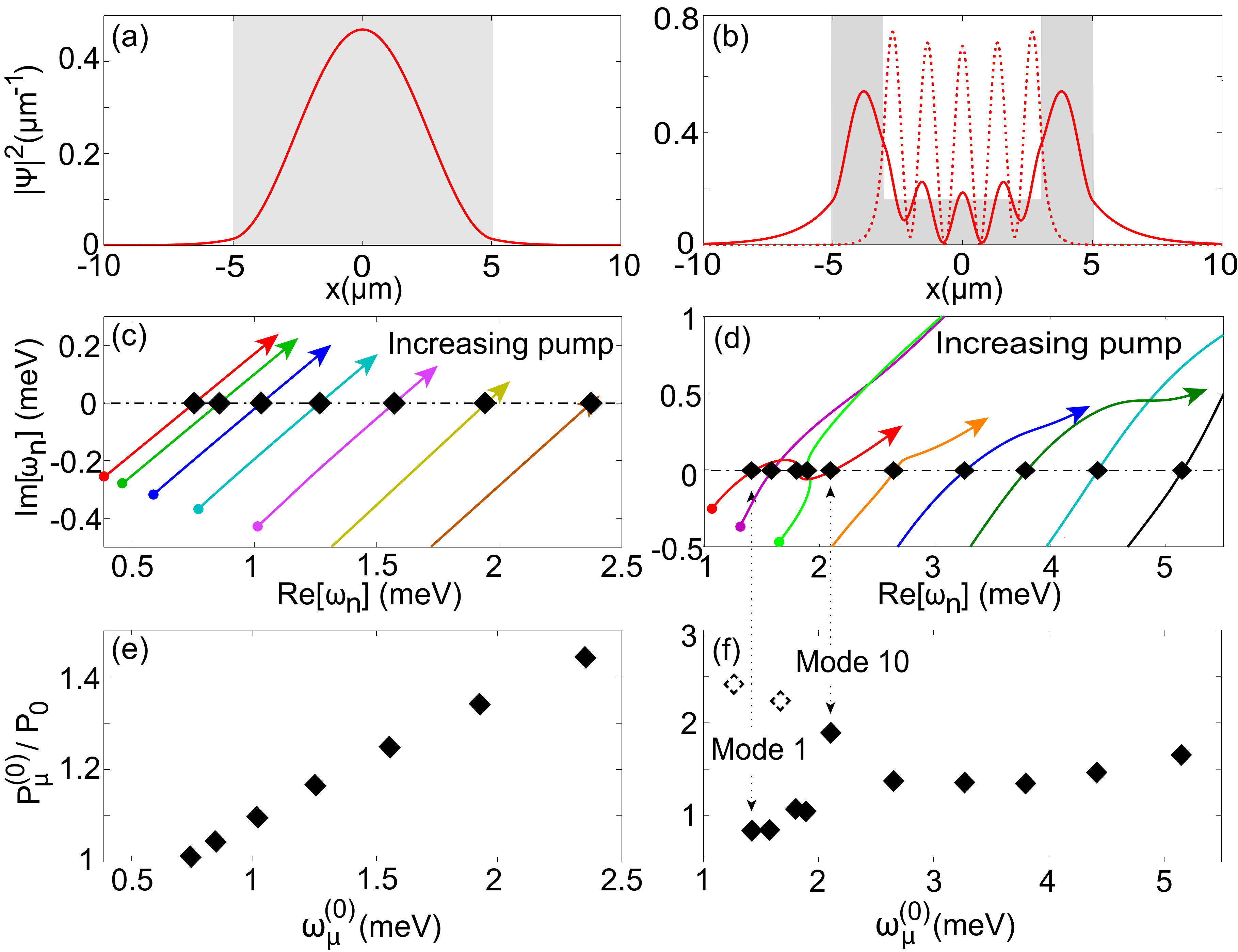}
\caption{(Color online) Exciton-polariton condensate in single-barrier (left column) and (right column) pump configurations. (a,b) First condensate pattern at the lowest threshold $P_1^{(0)}/P_0=1.01,0.84$ (solid line), respectively. Here $P_0=\gamma_c\gamma_{\!\scriptscriptstyle R}/R=100\, \text{meV\,$\mu$m$^{-2}$}$. Dashed line in (b) shows the non-isomorphically trapped mode 10 marked in (d) and (f). The shadowed area indicates the pump region in $|x|<5\,\mu\text{m}$. In (a) it is uniform and in (b) $f(x)$ in the basin is 1/5 its value in the barrier regions. (c,d) Trajectories (solid lines) of the complex frequencies $\omega_n$ for $P/P_0\in[0.5, 1.5]$ in (c) and $[0.5, 3.6]$ in (d). The start and end of each trajectory are indicated by the dot and the large arrowhead, respectively. Dash-dotted line indicates the real axis, on which the lowest seven linear thresholds in the pump configuration (a) and ten in (b) are marked by black diamonds. These threshold values are plotted against the corresponding frequency in (e,f). Open diamonds in (f) show two non-isomorphically trapped patterns similar to mode 10. Their trajectories are not shown in (d) for clarity.
Parameters used are similar to those in Refs.~\cite{Wouter_prb08, vortex}: $\gamma_c=1 \,\text{meV}$, $\gamma_{\!\scriptscriptstyle R}=10\,\text{meV}$, $m^{-1}=0.59\,\text{$\mu$m}^2\,\text{meV}$, $g_{\!\scriptscriptstyle R} = 0.072\,\text{$\mu$m}^2\,\text{meV}$, and $R = 0.1\,\text{$\mu$m}^2\,\text{meV}$.
}\label{fig:1D}
\end{center}
\end{figure}

Similar to the pole-pulling picture of how a laser reaches its threshold \cite{Haken,pra10}, the linear thresholds here can be found by studying the pump-dependent motions of the resonances $q_n$ of the light-induced potential, defined by
\be
\left[-\nabla^2 + s P f(\vec{r})\right]\phi_n(\vec{r}; P) = q_n^2(P) \phi_n(\vec{r};P)\label{eq:GPE1}
\ee
with an outgoing boundary condition outside the pump region and $s\equiv{m}(2g_{\!\scriptscriptstyle R} + i R)/{\gamma_{\!\scriptscriptstyle R}}$. The linear thresholds are reached when $\omega_n\equiv q_n^2/2m-i\gamma_c/2$ becomes real [Fig.~\ref{fig:1D}(c),(d)], at which $\omega_n$ gives the frequency of the corresponding condensate pattern in Eq.~(\ref{eq:GPE0}) when all other $\psi_\mu(\vec{r})\approx 0$. We denote these linear threshold frequencies by $\omega_\mu^{(0)}$ [Fig.~\ref{fig:1D}(e),(f)].

The linear thresholds $P^{(0)}_\mu$ and the corresponding condensate patterns $\psi_\mu(\vec{r})$ satisfy a continuity relation
\be
\left[\gamma_c-\frac{RP^{(0)}_\mu f(\vec{r})}{\gamma_{\!\scriptscriptstyle R}}\right] \rho_\mu(\vec{r}) + \nabla\cdot \vec{j}_\mu(\vec{r})=0, \label{eq:current}
\ee
where $\rho_\mu(\vec{r})\equiv|\psi_\mu(\vec{r})|^2$ is the probability density of the condensate and $\vec{j}_\mu(\vec{r})\equiv\frac{i}{2m}[\psi_\mu(\vec{r})\nabla\psi^*_\mu(\vec{r}) - c.\,c.]$ is the probability current \cite{Keeling,Wouter_prb08} within the condensate plane.
By integrating over the pump region, we find a simple expression for the linear thresholds
\be
P^{(0)}_\mu = \frac{\gamma_\mu + \gamma_c}{\frac{R}{\gamma_{\!\scriptscriptstyle R}}\mathcal{G}_\mu}.\label{eq:threshold}
\ee
Its numerator gives the total loss from the light-induced potential, including the out-of-plane mirror loss $\gamma_c$ and the in-plane leakage $\gamma_\mu \equiv \oint_\text{pump} \vec{j}_\mu(\vec{r})\cdot d\vec{s}$. Its denominator is proportional to $\mathcal{G}_\mu \equiv \int_\text{pump} f(\vec{r})\rho_\mu(\vec{r}) d\vec{r}$, which is a measure of how strongly the condensate overlaps with the pump spatially; the stronger the overlap, the lower the threshold. Note that we have used $\int_\text{pump} \,\rho_\mu(\vec{r}) d\vec{r}=1$. When the pump is translationally invariant, no current flows and the local gain balances the mirror loss, with which we recover the familiar result $P_0=\gamma_c\gamma_{\!\scriptscriptstyle R}/R$ \cite{Wouter_prb08}.

In Fig.~\ref{fig:1D} we show two one-dimensional (1D) examples with single-barrier and double-barrier pump configurations.
In the first case all $\mathcal{G}_\mu=1$, indicating that the condensate patterns overlap equally with the pump. The linear thresholds in this case are simply given by $P^{(0)}_\mu/P_0 = 1 + \gamma_\mu/\gamma_c\gtrsim 1$. The fundamental mode with a single peak [Fig.~\ref{fig:1D}(a)] has the smallest in-plane leakage $\gamma_\mu$ and thus the lowest linear threshold [Fig.~\ref{fig:1D}(e)]. This is not the case in the double-barrier configuration. Both the in-plane leakage $\gamma_\mu$ and pump overlap $\mathcal{G}_\mu$ exhibit similar non-monotonic behaviors in the low frequency region [$\omega_\mu^{(0)}<2.5$~meV in Fig.~\ref{fig:1D}(f)]. In fact these condensate patterns can be divided into two categories: one resides mostly in the basin, which has a small in-plane leakage but does not overlap strongly with the pump [see mode 10 represented by the dashed line in Fig.~\ref{fig:1D}(b)]; the other resides primarily within the two barriers, benefitting from the exciton reservoir but has a stronger in-plane leakage [see mode 1 represented by the solid line in Fig.~\ref{fig:1D}(b)]. We refer to these two categories as {\it non-isomorphically} and {\it isomorphically} trapped patterns, depending on whether the condensate pattern follows the geometry of the pump. In the double-barrier configuration it is the isomorphically trapped mode 1 that has the lowest threshold, which is even lower than $P_0$ thanks to a tiny $\gamma_\mu/\gamma_c=0.167$ and a large $\mathcal{G}_\mu=1.39$. We note that a non-isomorphically trapped pattern and an isomorphically trapped one can result from the same linear resonance. Modes 1 and 10 discussed above is representative of such a case, which are on the same trajectory in Fig.~\ref{fig:1D}(d).

\begin{figure}[t]
\begin{center}
\includegraphics[clip,width=0.95\linewidth]{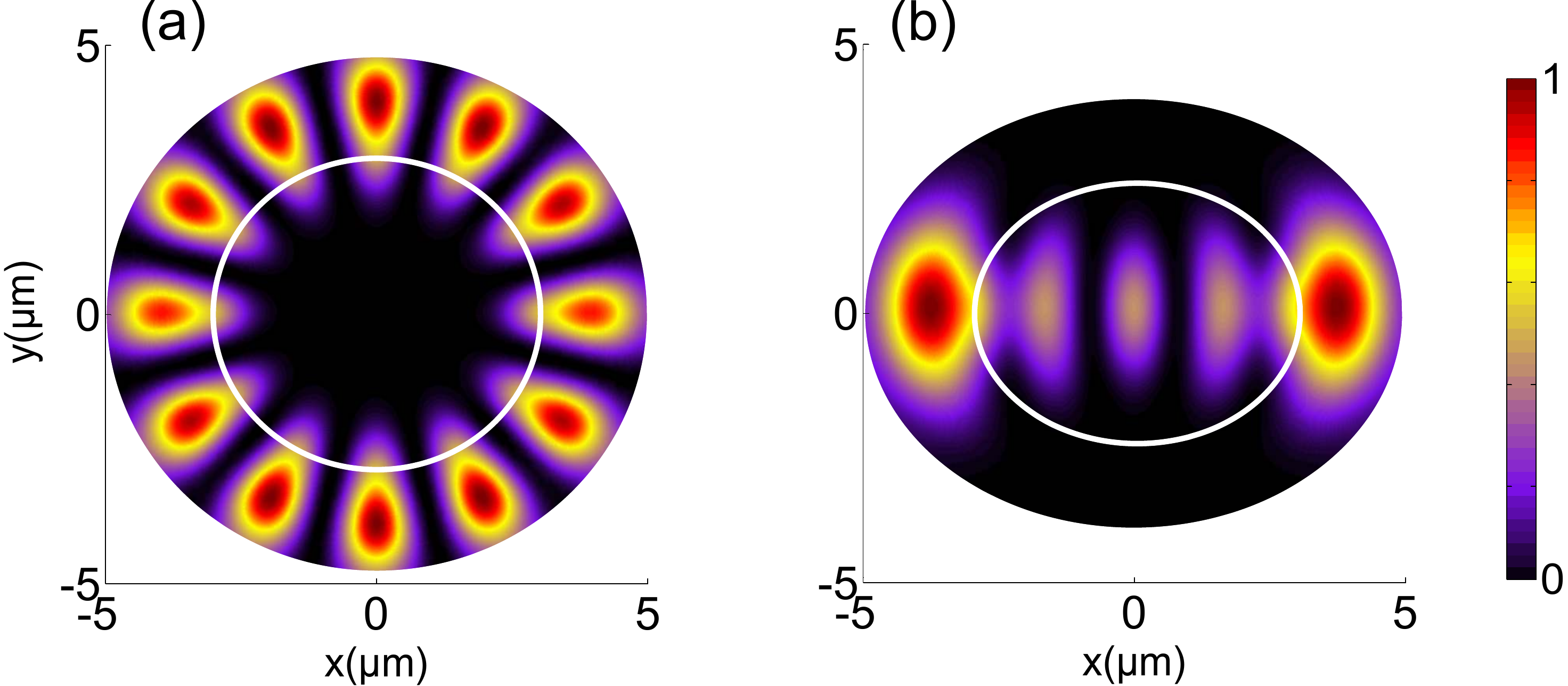}
\caption{(Color online) Intensity pattern of the first condensate mode with an elliptical ring pump. The eccentricity is $e=0.4,0.6$, respectively. Their pump profile along $y=0$ is identical with Fig.~\ref{fig:1D}(b). The
white solid line indicates the inside boundary of the ring pump. Other parameters used are the same as in Fig.~\ref{fig:1D}.
} \label{fig:2D}
\end{center}
\end{figure}

The lowest threshold pattern can change from one category to the other upon a slight perturbation of the pump in some cases. This is illustrated in the two-dimensional (2D) example shown in Fig.~\ref{fig:2D} with an elliptical ring pump. If the eccentricity $e<0.5$, the lowest threshold pattern is an isomorphically trapped whispering-gallery mode with a dominant angular momentum number $M=6$ [see Fig.~\ref{fig:2D}(a)], similar to the experimental observation reported in Ref.~\cite{Manni}.
The situation changes as we further deform the ring pump. As Fig.~\ref{fig:2D}(b) shows, at $e=0.6$ the condensate pattern of the lowest threshold is non-isomorphically trapped, with a significant fraction outside the ring region and similar to what has been observed in Ref.~\cite{Baumberg2}. This transition again highlights the conflicting roles of the pump profile: the in-plane leakage of this non-isomorphically trapped mode is lower than the mirror loss ($\gamma_1/\gamma_c=0.54$), while that of the second mode similar to the pattern shown in Fig.~\ref{fig:2D}(a) is not ($\gamma_2/\gamma_c=1.41$). This better confinement helps achieving a lower threshold, despite its weaker overlap with the pump, i.e. $\mathcal{G}_1=1.68$ versus $\mathcal{G}_2=2.54$.

With the nonlinear interactions taken into consideration, the only modification to the continuity relation (\ref{eq:current}) comes from the gain saturation, i.e.
\be
\left[\gamma_c-\frac{R P f(\vec{r})}{\gamma_{\!\scriptscriptstyle R} + R\varrho(\vec{r})}\right] \rho_\mu(\vec{r}) + \nabla\cdot j_\mu(\vec{r})=0, \label{eq:currentNon}
\ee
in which $\varrho(\vec{r}) = \sum_\nu|\psi_\nu(\vec{r})|^2$ and we have neglected the beating terms between different condensate patterns
in the multimode regime.
We note that the direct polariton-polariton interaction is real-valued and does not enter the above expression. Nevertheless, it underlines a major difference from photon lasers, with the nonlinearity occurring both inside and outside the effective cavity formed by the light-induced potential. Therefore, the boundary of the system we consider in the nonlinear regime needs to be beyond the spatial extension of the condensates, at which the polariton density is negligible; in the linear analysis the system boundary is effectively set by the pump spot, beyond which the condensate profile is determined by the (outgoing) continuity conditions.

To determine the nonlinear condensate patterns, we expand them in the GPE using a linear basis $\{\varphi_n(\vec{r};\omega)\}$, parameterized by a real-valued frequency $\omega$:
\be
\left[\nabla^2 + q^2\right]\varphi_n(\vec{r};\omega) = sf(\vec{r})P_n \varphi_n(\vec{r};\omega), \label{eq:GPE2}
\ee
$P_n$ are the generalized eigenvalues and $q^2 \equiv m(2\omega+i{\gamma_c})$. $\varphi_n(\vec{r};\omega)$ are non-Hermitian due to the openness, and they satisfy the following biorthogonal relation
\be
\int d\vec{r} f(\vec{r}) \phi_n(\vec{r};\omega)\phi_j(\vec{r};\omega) \equiv \bracket{n|f(\vec{r})|j} = \delta_{mn},
\ee
instead of the more familiar power orthogonality relation. We note that all basis functions have the same decay rate $\gamma_c$ outside the pump region.

For each condensate mode $\mu$, we choose $\omega$ to be identical to $\omega_\mu$, the yet to be determined condensate frequency, and write $\psi_\mu(\vec{r}) = \sum_n a^\mu_n \varphi_n(\vec{r};\omega_\mu)$.
Henceforth the $\omega_\mu$-dependence of $\{\varphi_n(\vec{r};\omega_\mu)\}$ is suppressed, and the GPE takes the following self-consistent form:
\be
\sum_j \mathcal{T}_{nj}a_j^\mu ={sP_n} a_n^\mu,\,\, \mathcal{T}_{nj} = \bracket{n|[S(\vec{r})f(\vec{r})P + 2mg\varrho(\vec{r})]|j}, \nonumber
\ee
where $S(\vec{r})\equiv m(2g_{\!\scriptscriptstyle R}+i{R})/(\gamma_{\!\scriptscriptstyle R} + R\varrho(\vec{r}))$. The above derivation parallels its counterpart for a photon laser, the Steady-state Ab-initio Laser Theory (SALT) \cite{pra07,Science,pra10}.

\begin{figure}[t]
\begin{center}
\includegraphics[clip,width=\linewidth]{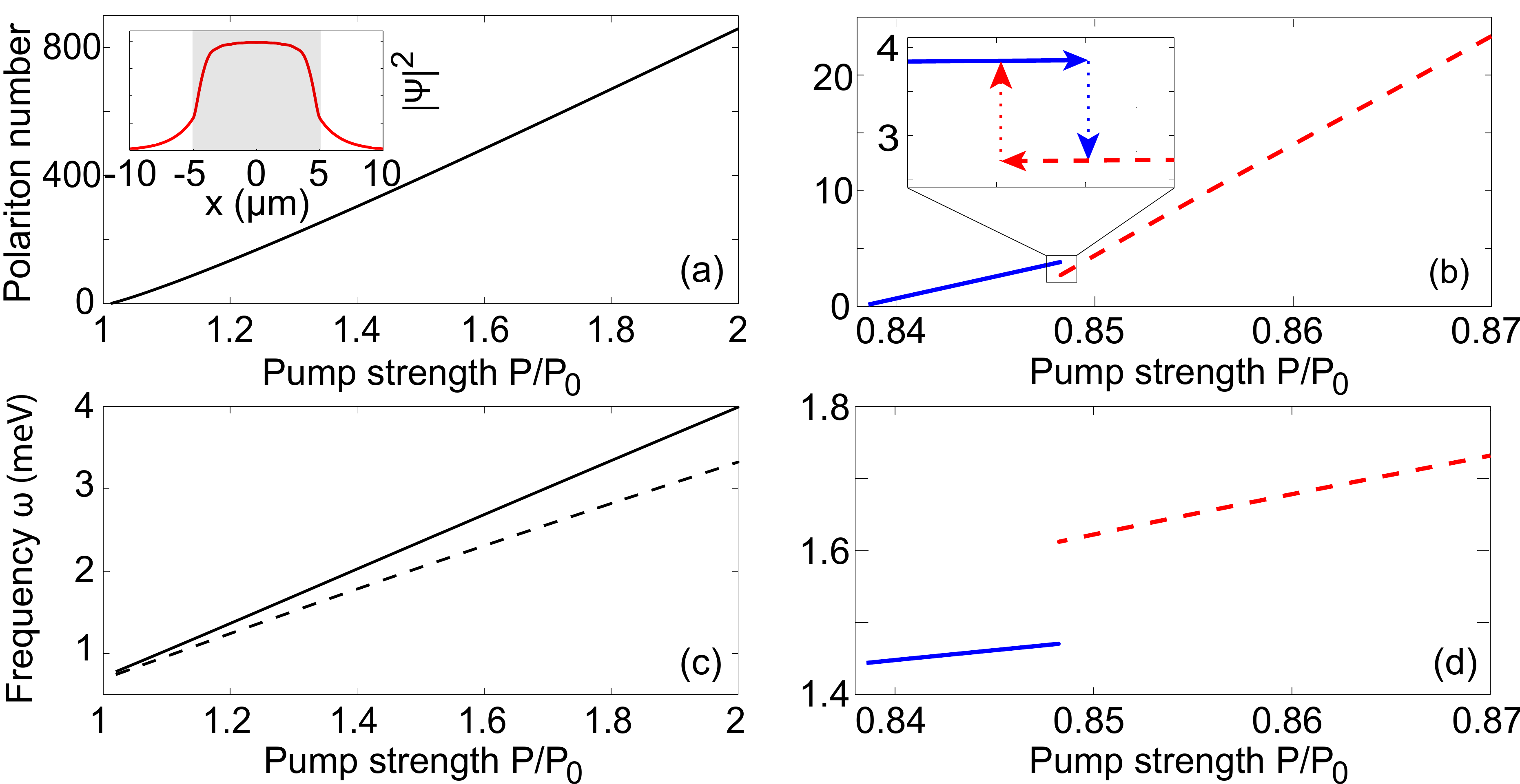}
\caption{(Color online) (a,b) Polariton number in each condensate pattern in the single-barrier (left) and double-barrier (right) pump configurations discussed in Fig.~\ref{fig:1D}. The inset in (a) shows the less localized condensate pattern at $P/P_0=1.5$. The inset in (b) zooms onto the bistable region $P/P_0\in[0.84825,0.84826]$. The corresponding condensate frequencies are shown in (c,d). The dashed line in (c) shows the condensate frequency without considering the kinetic energy term. $g=0.04\,\text{$\mu$m}^2\,\text{meV}$.} \label{fig:1D_non}
\end{center}
\end{figure}

It is well known that the condensate frequency shifts with the pump strength, which is attributed to the polariton-polariton interaction, as well as the saturated polariton-exciton reservoir interaction \cite{Savvidis,Kavokin,Wouter_prl07,Ferrier_PRL106}.
With a spatially inhomogeneous pump, however, there is another term that contributes to the frequency shift, namely the change of the kinetic energy ${\omega}_\text{kin}(P) \approx \int  {|\nabla w(\vec{r})|^2}/{2m}\, d\vec{r}$,
where $w(\vec{r}) = \psi(\vec{r})/\sqrt{N}$ is the normalized wave function of the condensate and $N=\int \varrho(\vec{r}) d\vec{r}$ is the total number of condensed polaritons. Unlike the other two contributions, it does not have an explicit dependence on the pump strength or the polariton number. Nevertheless, it still depends on the pump strength strongly. Take the 1D single-barrier pump configuration for example. The fundamental mode
fully suppresses all the higher-order patterns even at $P=2P_0$. This can be explained by the pump dependent pattern reconfiguration: the lowest threshold mode  [Fig.~\ref{fig:1D}(a)] becomes {\it less localized} to take better advantage of the pump [Fig.~\ref{fig:1D_non}(a)]. This spatial reconfiguration results in a kinetic blueshift of $0.63\, \text{meV}$ at $P=2P_0$, which is about $21\%$ of the total [Fig.~\ref{fig:1D_non}(c)].

\begin{figure}[t]
\begin{center}
\includegraphics[clip,width=0.95\linewidth]{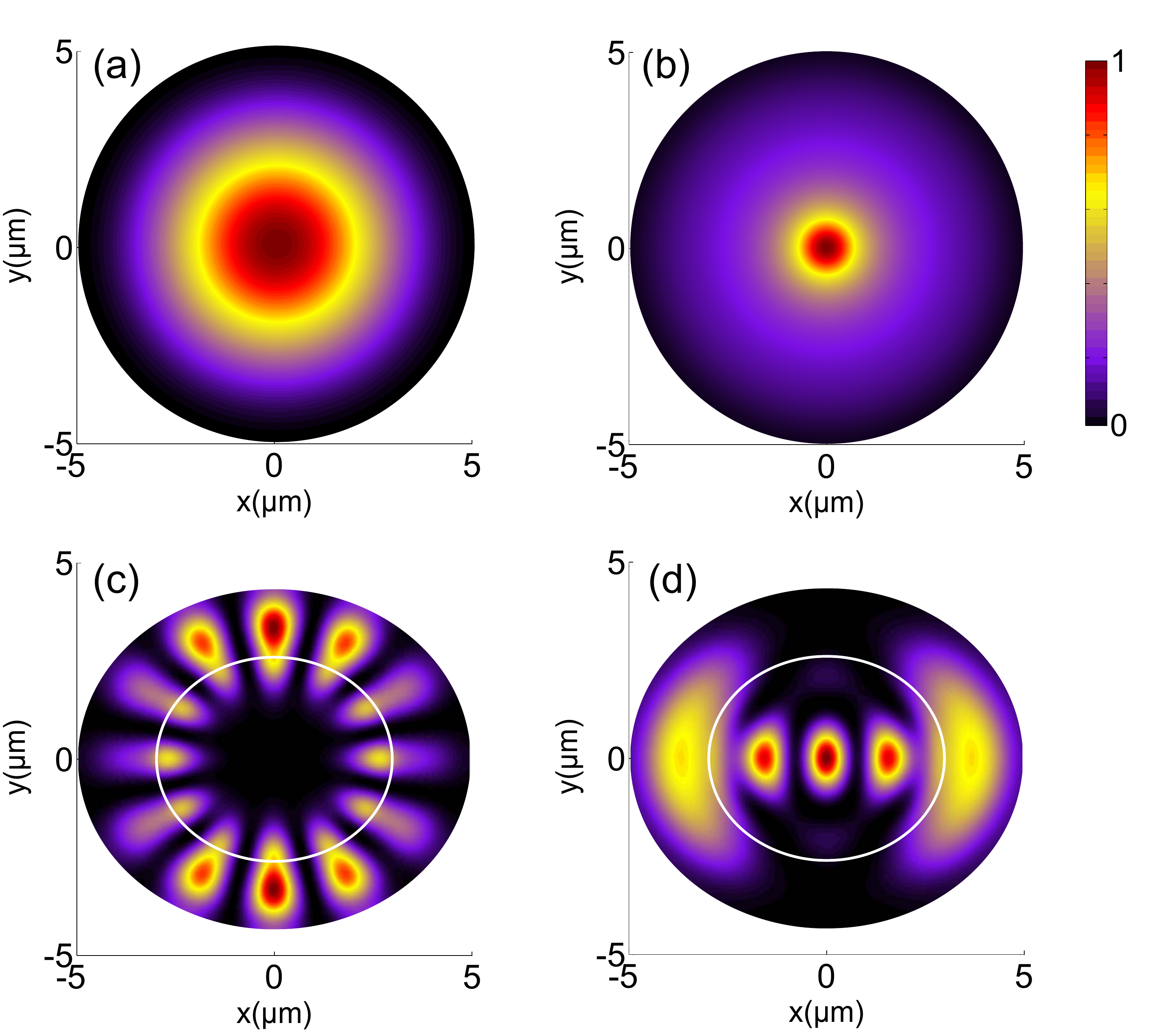}
\caption{(Color online) (a,b) Spatial pattern of the single-mode condensate in a uniform disk pump of radius $r=5\mu{m}$, calculated at $P=P^{(0)}_1=1.03P_0$ and $1.4P_0$, respectively. (c,d) Cooperative pattern reconfiguration of the first two patterns in an elliptical ring pump of eccentricity $e=0.5$, calculated at $P=1.1P_0$. They oscillate in the condensate simultaneously.} \label{fig:2D_non}
\end{center}
\end{figure}

The evolution of the condensate pattern is strongly geometry dependent. For a uniform disk pump in 2D, the fundamental mode of angular momentum $M=0$ [Fig.~\ref{fig:2D_non}(a)] is also the condensate pattern with the lowest threshold and suppresses all the higher-order ones even at $P=2P^{(0)}_1$. As the pump strength increases, it becomes {\it more localized} at the origin [Fig.~\ref{fig:2D_non}(b)], which is in stark contrast to the 1D single-barrier case. The enhanced localization is due to the mixing with higher-order modes $\varphi_n(\vec{r})$ of $M=0$, which are more localized.

The nonlinear effects become more dramatic in complex pump configurations. For the double-barrier pump configuration for example, we find that interaction-induced mode switching occurs frequently. As Fig.~\ref{fig:1D_non}(b,d) shows, a discontinuity in both the number of condensed polaritons and the frequency of the condensate takes place near $P=0.848P_0$. If we zoom in to this discontinuity, a narrow region of bistability \cite{bistability1,bistability2,bistability3} can be identified, where these two patterns cannot form simultaneously.
Detailed stability analysis will be given elsewhere \cite{stability}.

Multi-mode condensates have been observed previously \cite{Baas, Bloch1,Baumberg1}. We find this to be the case for the elliptical ring pump discussed above. Interestingly, a cooperative multi-pattern reconfiguration takes place between the isomorphically and non-isomorphically trapped patterns. For an eccentricity $e=0.5$, the two condensate modes with the lowest thresholds are similar to those shown in Fig.~\ref{fig:2D}, with the isomorphically trapped pattern having a slightly lower threshold than the other.
They coexist beyond $P\approx P_0$, and their nonlinear interaction re-sculptures their patterns, with the second mode becoming even more non-isomorphically trapped [Fig.~\ref{fig:2D_non}(d)]. The isomorphically trapped pattern on the other hand, has its brightest spots along the minor axis further enhanced, and reduces its spatial overlap with the other [Fig.~\ref{fig:2D_non}(c)] for a symbiotic utilization of the pump by both patterns.

In summary, we have shown that the threshold of pattern formation in exciton-polariton condensates can be understood intuitively and quantitatively through a continuity relation. For spatially complex pump configurations, the delicate balance between particle formation and leakage can result in condensates that are either isomorphically or non-isomorphically trapped. We have also discussed the rich nonlinear behaviors including single- and multi-pattern reconfiguration and interaction-induced mode switching.

We acknowledge support by DARPA Grant No. N66001-11-1-4162 and NSF CAREER Grant No. DMR-1151810.


\begin{thebibliography}{99}
\bibitem{bibnote:1} Two exceptions are random lasers in the weakly scattering regime which do not possess any well-defined resonances and wide-area VCSELs.
\end{thebibliography}
\end{document}